\documentclass[preprint,prc,showpacs,preprintnumbers,amsmath,amssymb,floatfix]{revtex4}
\usepackage{amsmath}
\usepackage{graphicx}
\usepackage{dcolumn}
\usepackage{bm}
\usepackage{ulem} 
\usepackage[usenames]{color}
\usepackage{epstopdf}
\usepackage{epsfig}
\usepackage{float}
\usepackage{subfigure}
\usepackage{multirow}
\usepackage[version=3]{mhchem}
\usepackage{tikz}

\newcommand{\nc}{\newcommand}       
\nc{\vc}[1] {\mbox{\boldmath $#1$}} 
\nc{\del}       {\partial}              
\nc{\bra}       {\langle}               
\nc{\ket}       {\rangle}               
\nc{\bras}[1]   {\langle #1|}           
\nc{\kets}[1]   {|#1\rangle}            
\nc{\mapleft}[1]{           
	\smash{\mathop{\,          %
			\hbox to 1.5cm{\rightarrowfill}\, }\limits_{#1}}}
\nc{\beq}     {\begin{eqnarray}} \nc{\eeq}    {\end{eqnarray}}
\nc{\nn}      {\\\nonumber} \nc{\vs}      {\vspace{-0.275cm}}
\nc{\fra}    {\frac{1}{2}}
\nc{\mb}        {\mathbf}

\begin{document}
	
	
	\title{The one-pion-exchange potential with contact terms from lattice QCD simulations}
	
	\author{Jinniu Hu}
	\email{hujinniu@nankai.edu.cn}
	\affiliation{School of Physics, Nankai University, Tianjin 300071, China}
	\author{Ying Zhang}
	\email{yzhangjcnp@tju.edu.cn}
	\affiliation{Department of Physics, Faculty of Science, Tianjin University, Tianjin 300072, China}
	\author{Hong Shen}
	\email{songtc@nankai.edu.cn}
    \affiliation{School of Physics, Nankai University, Tianjin 300071, China}
    \author{Hiroshi Toki}
    \email{toki@rcnp.osaka-u.ac.jp}
    \affiliation{Research Center for Nuclear Physics (RCNP), Osaka University, Ibaraki, Osaka 567-0047, Japan}

	\date{\today}
	
\begin{abstract}

	The pion-mass-dependent nucleon-nucleon ($NN$) potentials {in term of one-pion exchange and contact terms}  are obtained from the latest lattice QCD simulations of two-nucleon system, {which employ the forms of leading order (LO) $NN$ potential from the chiral effective field theory and thus are named as the LO chiral potential in this work}. We extract the coefficients of contact terms and cut-off momenta in these potentials, for the first time, by fitting the phase shifts of $^1S_0$ and $^3S_1$ channels generated by the results of HALQCD collaboration with various pion masses from $468.6$ MeV to $1170.9$ MeV.  The low-energy constants in the $^1S_0$ and $^3S_1$ channels become weaker and approach each other for larger pion masses.  These LO chiral potentials are applied to symmetric nuclear and pure neutron matter within the Brueckner-Hartree-Fock method.   At this moment, however, we do not have yet the information of the $P$-wave $NN$ interaction to be provided by the lattice QCD simulations for full description of nuclear matter.  Our results will enhance the development of nuclear structure and nuclear matter by controlling the contribution of the pionic effect and illuminate the role of chiral symmetry of the strong interaction in complex system.
	
\end{abstract}
	
	
	
	\maketitle
	
\section{Introduction}
Last forty years, the chiral effective field theory (ChEFT) and lattice quantum chromodynamics (LQCD) are two powerful tools to solve the problems of strong interaction at the low energy scale due to its nonpertubative behaviors. The realization of LQCD simulation is strongly dependent on the computing resource. It is very hard to describe the properties of hadrons at physical pion mass, $m_\pi\approx 140$ MeV. On the other hand, these properties can be generated as functions of pion masses in ChEFT. Therefore, there are many attempts to extrapolate the LQCD results of hadrons to physical pion mass with the help of ChEFT, such as baryon masses, axial coupling constants, magnetic moments~\cite{akhan06,duerr14,ren17,xiao18}, and so on. These works largely improved our understanding on the QCD theory for single hadrons at low energy region.

For nucleon-nucleon ($NN$) system, the potentials from ChEFT at different expansion orders were constructed, whose lower-energy constants (LECs) were determined by fitting the $NN$ scattering data~\cite{weinberg90,weinberg91,weinberg92,ordonez94,ordonez96,epelbaum98,epelbaum00,entem03,epelbaum05}.  Until now, the next-to-next-to-next-to-next-to-leading order (N$^4$LO, fifth order)  has been completely included in the chiral $NN$ potentials~\cite{epelbaum15a,epelbaum15b,entem15,Entem:2015xwa,entem17,reinert17}, which were adopted to calculate the properties of light nuclei and infinite nuclear matter~\cite{epelbaum15b,hu17,binder18}. As for the LQCD, the information of nucleon-nucleon interaction were extracted from the Wilson quark action by several groups~\cite{ishii07,beane13,iritani17,berkowitz17}. The pion-mass-dependent $NN$ forces were obtained by the HALQCD collaboration in the LQCD~\cite{aoki10,aoki12,inoue11,inoue12}. Although, the pion masses (quark masses) in these calculations were still far from the physical mass point ($m_\pi=468.6, ..., 1170.9$ MeV), they were already applied to investigate the properties of nuclear many-body problems~\cite{inoue13,inoue15,mcilroy18,hu16}.  

As the case of ChEFT for single nucleon at various pion masses provided important information on the chiral dynamics, the ChEFT for two-nucleon system should provide valuable information on the role of pion in complex nuclear system.   However, the investigations for the connection of the ChEFT and LQCD for nuclear force is very few until now.  Recently, Song {\it et al.} adopted the available hyperon-hyperon scattering data from LQCD simulations to discuss the relativistic baryon-baryon interaction~\cite{song18}.  Barnea {\it et al.} predicted the binding energies of light nuclei with the $NN$ interaction generated by the pionless EFT using the hyperspherical harmonics method and auxiliary-field diffusion Monte Carlo method at $m_\pi=805$ MeV, where the leading-order (LO) LECs were fitted to the deuteron, dineutron, and triton energies obtained by the NPLQCD group~\cite{barnea15, contessi17}. 

In this work, we try to construct a bridge between ChEFT and LQCD in terms of $NN$ force and extract the LECs from LQCD simulations for the first time. {Though it is not any more available for the ChEFT at larger pion mass,} it is still important to extract the pion-mass dependence of LECs to elucidate the role of pion in complex system from LQCD, which can change the pion mass freely.  The chiral $NN$ potentials are generated from the available HALQCD lattice simulations, which have strong pion-mass dependence. Since the statistical errors of the lattice simulations are not satisfactory yet, the chiral expansion order is restricted to the lowest order, i.e., leading order, where only two LECs, $C_S$ and $C_T$ can be extracted.  We will discuss the pion-mass dependence of LECs and the pion form factors. Furthermore, we will apply these chiral potentials to study the properties of nuclear matter and obtain their equations of state with different pion masses.

\section{Leading order chiral $NN$ potential}
Since the chiral effective Lagrangian at  leading order comes from two-nucleon tree diagrams, the $NN$ potential is written as one-pion-exchange component and the contact terms. It is given explicitly as following in the center-of-mass system~\cite{epelbaum09, machleidt11},
\beq
V^{(0)}_{NN}(\vec p_1,\vec p_2)=-\frac{g^2_A}{4f^2_\pi}\frac{\vec\sigma_1\cdot \vec q\vec\sigma_2\cdot\vec q}{\vec q^2+m^2_\pi}\vec\tau_1\cdot\vec\tau_2+C_S+C_T\vec\sigma_1\cdot\vec\sigma_2,
\eeq  
where $g_A,~f_\pi$ and $m_\pi$ denote the nucleon axial coupling constant, pion decay constant, and pion mass, respectively. $\vec q=\vec p_1-\vec p_2$ is the momentum transfer between two nucleons, where $\vec p_1$ and $\vec p_2$ denote the initial and final nucleon momenta in the center-of-mass system, respectively. The superscript (0) denotes the chiral expansion order. The LECs in the LO contact terms can be expressed in terms of the partial wave components in the $^1S_0$ and $^3S_1$ channels~\cite{entem17},
\beq
V^{(0)}_\text{ct}(^1S_0)&=&\widetilde C_{^1S_0}=4\pi (C_S-3C_T),\nn
V^{(0)}_\text{ct}(^3S_1)&=&\widetilde C_{^3S_1}=4\pi (C_S+C_T).
\eeq
To determine the LECs, $\widetilde C_{^1S_0}$ and $\widetilde C_{^3S_1}$, the LO chiral potential,  $V^{(0)}_{NN}(\vec p_1,\vec p_2)$, is inserted to a scattering equation to generate the phase shifts of $NN$ collisions. We follow the choice of Entem {\it et al.} with a relativistic version, the Blankenbeclar-Sugar equation, which is shown explicitly in~\cite{brockmann90,entem17}. Furthermore, a regular function should be multiplied with the chiral potential, $\widehat V$, to avoid the divergence at high momenta in the scattering equation, with an exponential function in momentum space~\cite{entem17,reinert17},
\beq
f(p_1,p_2)=\exp\left[-(p_1/\Lambda)^{(2n)}-(p_2/\Lambda)^{(2n)}\right].
\eeq
The cut-off momentum, $\Lambda$,  and the power, $n$, will be fixed by fitting the phase shifts from the lattice forces in this work. Once these chiral $NN$ potentials are obtained, the properties of symmetric nuclear matter and pure neutron matter will be calculated within Brueckner-Hartree-Fock (BHF) method~\cite{li06,baldo07,baldo12}.

\section{Results and discussions}
In the lattice $NN$ forces from HALQCD collaboration, the pion masses range from $468.6$ MeV to $1170.9$ MeV with different $u,~d$ quark masses. The corresponding nucleon masses change from $1161$ MeV to $2274$ MeV~\cite{inoue11,inoue13}. Furthermore, the axial-vector coupling constant $g_A$ and pion-decay constant $f_\pi$ can be evaluated by the LQCD method as a function of pion mass. The available investigations indicated that the axial-vector coupling constant is not so sensitive to the pion mass~\cite{akhan06}. Therefore, it is fixed to $g_A=1.20$ in the present work for all potentials as shown in Ref.~\cite{akhan06}. On the contrary, the pion-decay constant is strongly dependent on the pion mass from the present  LQCD simulations. Its value can be compared with the predictions of SU(2) chiral perturbation theory at next-to-next-to-leading order, which can be expressed by the chiral logarithms in terms of pion masses~\cite{duerr14}. Therefore, we can use these achievements to fix the pion-decay constants  at the corresponding pion masses from  $468.6$ MeV to $1170.9$ MeV. 

Once these physical constants are determined, there are four free parameters in the LO chiral $NN$ potentials, two LECs in contact terms, $ \widetilde C_{^1S_0}, ~\widetilde C_{^3S_1}$, the power of cut-off, $n$, and cut-off momenta $\Lambda$ at each pion mass. Here, we want to state that $\Lambda$ in this work is just a cut-off momenta in the form factor, rather than the breaking scale in the conventional ChEFT. They will be fitted with the least square method to the phase shifts in the $^1S_0$ and $^3S_1$ channels for neutron-proton system from the HALQCD lattice $NN$ potentials, below the laboratory energy $E_\text{lab}=150.0$ MeV.  The best fitting results are obtained for the exponential cutoff function at $n=1$, and we first fix $n$ at this value and simulate the phase shifts of HALQCD $NN$ potentials at long distance.  All physical constants and free parameters, which are pion-mass dependent, are listed in Table ~\ref{const}. All the pion-decay constants $f_\pi$, LECs of contact terms {(in unit of $10^4$ GeV$^{-2}$ )}, and cut-off momenta increase with the pion masses. These LO chiral potentials are named as LO469, LO672, LO837, LO1015, and LO1171, corresponding to the different pion masses, $m_\pi=468.6,~672.3,~836.5,~1015.2$, and $1170.9$ MeV, respectively for convenience.

\begin{table*}[htb]
	\centering
	\begin{tabular}{r c c c c c}
		\hline
		\hline
		&    LO469~~~             &    LO672~~~             &    LO837~~~        &    LO1015~~~               &    LO1171~~~     \\
		\hline
		$m_\pi$ [MeV]~~~                      &  $468.6$~~~            &     $672.3$~~~        &  $836.5$~~~       &   $1015.2$~~~              &    $1170.9$      \\
		$M_N$  [MeV]~~~                       &  $1161.0$~~~           &     $1484.0$ ~~~     &  $1749.0$~~~      &   $2031.0$ ~~~           &   $2274.0$    \\		
		$f_\pi$ [MeV]~~~                       &  $118.0$~~~             &    $133.0$ ~~~        &   $145.0$~~~       &   $159.0$~~~               &   $172.0$           \\	
		\hline	                  	
		$\widetilde C_{^1S_0}$~~~  & $-0.077$~~~    & $-0.027$~~~   & $-0.024$~~~ &   $-0.021$ ~~~   &   $-0.016$    \\		
		$\widetilde C_{^3S_1}$~~~  & $-0.091$~~~    & $-0.029$ ~~~  & $-0.028$~~~ &   $-0.025$~~~    &   $-0.022$    \\	              	
		$\Lambda$ [MeV]~~~               &  $334.37$~~~          &  $550.00$ ~~~        & $569.81$~~~       &   $622.93$ ~~~         &   $664.91$    \\		
		\hline
		$\chi^2/d.o.f$~~~  & $9.385$~~~    & $8.805$~~~   & $0.389$~~~ &   $2.665$ ~~~   &   $0.928$    \\		
		\hline
		\hline
	\end{tabular}
	\caption{The pion mass, nucleon mass, pion-decay constant, LECs of contact terms and cut-off momenta at LO chiral potential from LQCD data with different pion masses from $468.6$ MeV to $1170.9$ MeV. {The LO LECs are in unit of $10^4$ GeV$^{-2}$.} }\label{const}
\end{table*}

\begin{figure}[htb] 
	\centering
	\includegraphics[width=18cm]{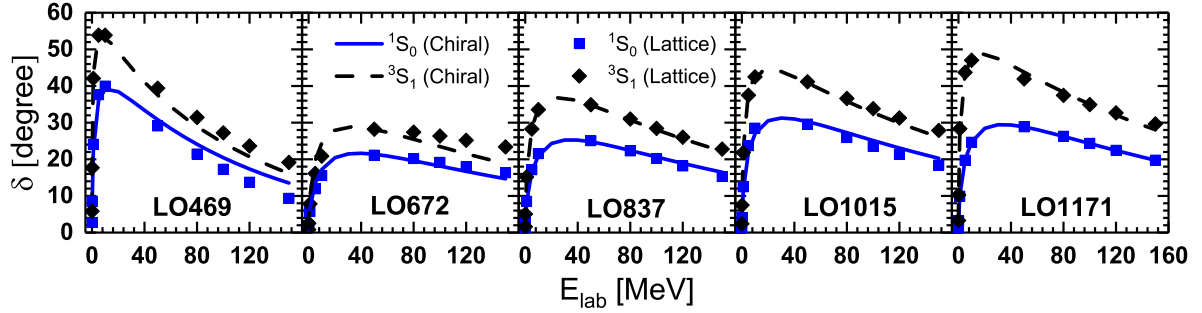}
	\caption{The phase shifts in the $^{1}S_0$ and $^{3}S_1$ channels from the lattice $NN$ potentials (discretized symbols) and the chiral $NN$ potentials (solid and dashed lines) with different pion masses from $468.6$ MeV to $1170.9$ MeV. }
	\label{ph}
\end{figure}

In Fig.~\ref{ph}, the $NN$ scattering phase shifts in the $^{1}S_0$ and $^{3}S_1$ channels from the HALQCD calculations and the LO chiral $NN$ potentials are compared at different pion masses. The solid and dashed curves are generated by the LO chiral $NN$ potentials. The square and diamond symbols represent the HALQCD data. It can be found that the LO chiral potentials can describe the results from lattice simulations very well at larger pion masses from LO837 to LO1171 potentials. This situation is becoming worse for lower pion masses, i.e., LO469 and LO672 potentials, at larger laboratory energy $E_\text{lab}$, which is related to the strong repulsions of $NN$ potential at short range region. It demonstrates that the contact terms of LO chiral potentials cannot completely display the complexities of lattice data at short range region and the next-to-leading order in the chiral expansion must be included at lower pion masses.	

\begin{figure}[htb] 
	\centering
	\includegraphics[scale=0.60]{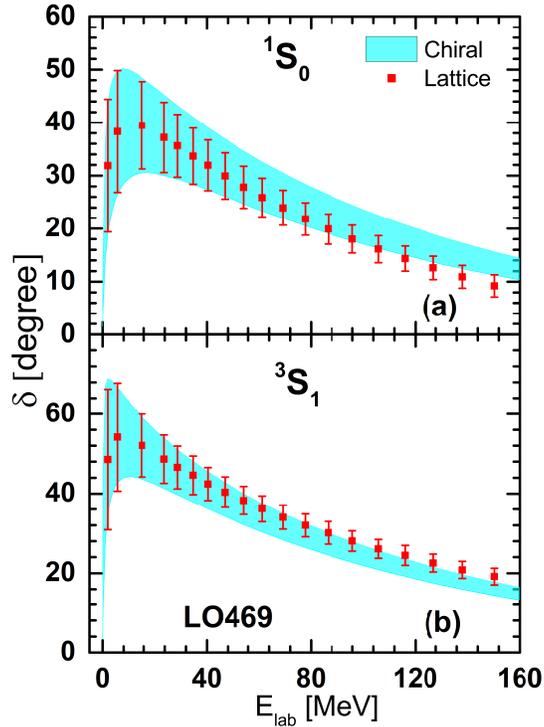}\\
	\caption{\label{error} The uncertainties of  phase shifts in the $^{1}S_0$ and $^{3}S_1$ channels from lattice $NN$ potential (square symbols with error bars) and the chiral $NN$ potential (shadow regions) with $m_\pi=468.6$ MeV. }
\end{figure}

The phase shifts from HALQCD collaboration involved stochastic and systematic errors due to the complicated Monte Carlo simulations, which have been shown in Fig.~4 of Ref.~\cite{inoue12} and Fig.~2 of Ref.~\cite{inoue13}. Therefore, it is necessary to discuss the influences of such uncertainties on the chiral potential, especially for the LECs, $ \widetilde C_{^1S_0}$ and $\widetilde C_{^3S_1}$. Therefore, these LECs are refitted by considering the error bars appearing in HALQCD phase shifts at $^1S_0$ and $^{3}S_1$ channels with $m_\pi=468.6$ MeV at a fixed cut-off momentum $\Lambda=334.37$ MeV. It is found that the corresponding LECs should be $\widetilde C_{^1S_0}=-0.077\pm0.012$ and $\widetilde C_{^3S_1}=-0.091\pm0.010$ now. The phase shifts from LO chiral potentials with such LECs are shown as the shadow regions in Fig.~\ref{error} for $^1S_0$ and $^{3}S_1$ channels in panels (a) and (b), respectively. The phase shifts from HALQCD with uncertainties are also given as the square symbols with error bars to be compared. It can be seen that the lattice data are within the LO chiral potentials' uncertainty regions for phase shifts at small energy region. {The differences between chiral potentials and HALQCD data become larger with increasing energy, because the cut-off momentum is relatively small in LO469 potentials. Now, the uncertainties of LECs in present chiral potentials are around $10\%\sim15\%$ from the LQCD data, while they are just about $1\%$ for the chiral potential generated by the $NN$ scattering data~\cite{reinert17}. Due to the lack of the uncertainties of HALQCD data at larger pion masses, a more systematical analysis cannot be done in present framework, such as the Bayesian method in Ref.~\cite{hu19}.  Furthermore, the calculations in HALQCD were done in the discretized space. However, present phase shifts are derived from the continuum momentum space. Therefore, the present uncertainties in LECs also include some finite-size effect.}

It is interesting to compare these chiral $NN$ potentials from LQCD data with the one at physical mass point. Therefore, a LO chiral potential is obtained by fitting the phase shifts at $^{1}S_0$ and $^{3}S_1$ channels extracted from the experimental data of $NN$ scattering  at $m_\pi=140$ MeV, where $M_N=939$ MeV and $f_\pi=92.4$ MeV, named as LO140 potential. The phase shifts are reproduced better than the one of the conventional LO chiral $NN$ potentials for the lower energy region $E_{lab}<150$ MeV~\cite{epelbaum15a}.  We use the order of the cutoff function $n=1$ instead of $2$ in the other works.  It means that the exponential regulator in this work is weaker than other ones, such as LO chiral potential given by Entem {\it et al.}~\cite{entem17}. Its cutoff momentum is $\Lambda=500$ MeV. Therefore, we can obtain more repulsive contributions on phase shifts.

\begin{figure}[htb] 
	\centering
	\includegraphics[scale=0.65]{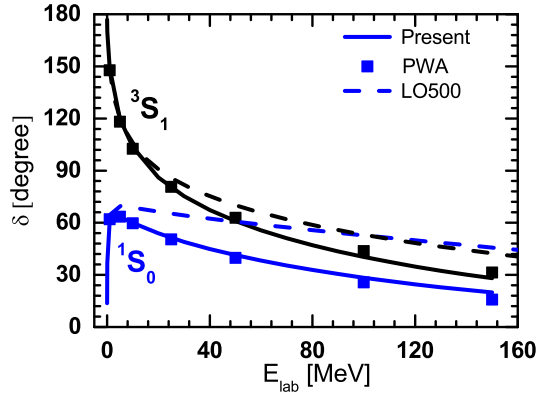}
	\caption{\label{ph140} The phase shifts of $^{1}S_0$ and $^{3}S_1$ channels from the Nijmegen PWA (discretized symbols), the chiral $NN$ potentials (solid lines) with physical pion mass $m_\pi=140.0$ MeV and small cut-off momentum in present work, and the chiral LO potential with $\Lambda=500$ MeV (dashed lines) given by Entem {\it et al.}~\cite{entem17}. }
\end{figure}

At the physical point, the strengths of contact terms and cut-off momenta are given as $\widetilde C_{^1S_0}= -0.152, ~\widetilde C_{^3S_1}=-0.214$, and $\Lambda=322.49$ MeV. With these parameters, the LO140 potentials at physical pion-mass point can describe the phase shifts from experimental data very well as shown in Fig.~\ref{ph140}, while the LO chiral potential with larger cutoff momentum shown as dashed line cannot reproduce the phase shifts at larger incident energies.   Comparing with the present lattice simulations at heavier pion masses, it is obvious that the phase shifts at $m_\pi=140$ MeV are larger in the $^{1}S_0$ and $^{3}S_1$ channels, which indicates there are more attractive contributions in the realistic $NN$ potentials. In the work of Inoue {\it et al.} of HALQCD group~\cite{inoue13}, the potentials of $^{1}S_0$ and $^{3}S_1$ channels only provide the maximum attractions around $-40$ MeV in the intermediate range, while these values are close to $-100$ MeV for realistic potentials, such as AV18  and CD Bonn potentials.   

\begin{figure}[htb] 
	\centering
	\includegraphics[scale=0.65]{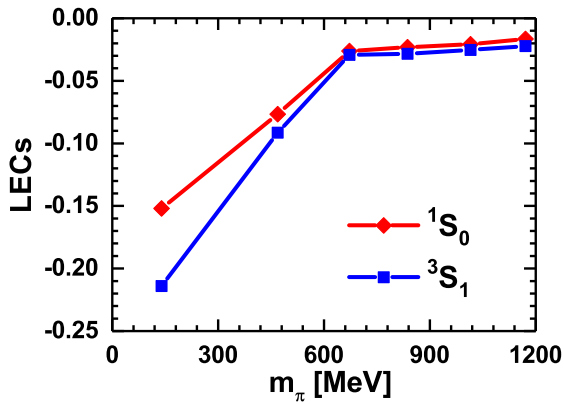}
	\caption{\label{ci} The LECs at leading order, $\widetilde{C}_{^1S_0}$ and $\widetilde{C}_{^3S_1}$ with different pion masses from $140.0$ MeV to $1170.9$ MeV. }
\end{figure}

In Fig.~\ref{ci}, the LECs at leading order, $\widetilde{C}_{^1S_0}$ and $\widetilde{C}_{^3S_1}$ are given as a function of pion mass from $m_\pi=140.0$ MeV to $m_\pi=1170.9$ MeV.  At larger pion masses, these LECs in the $^{1}S_0$ and $^{3}S_1$ channels are almost identical.  With the pion mass approaching the physical value, their differences become larger suddenly. {With present HALQCD data, it looks not so easy to extrapolate LECs of chiral potential at physical pion mass from the lattice data. Comparing to the LQCD simulations on meson and one-baryon sectors, the uncertainties of two-nucleon system from HALQCD is relatively larger. Furthermore, the mechanism of $NN$ interaction is more complicated. In this work, we just attempt to build the connections between the ChEFT and LQCD in $NN$ potentials. With more comprehensive LQCD calculations on few-body systems in future, it is expected that we can extract the $NN$ interaction at the physical pion mass using the lattice results at larger pion masses in the chiral expansion technique as has been performed for single nucleon.}

Recently there was an interesting study of two-nucleon system for axion production and gamma absorption assuming $\widetilde{C}_{^1S_0}=\widetilde{C}_{^3S_1}$, which corresponds the Wigner symmetry~\cite{mehen99}.  The Wigner symmetry is obtained in the large number of colors limit of QCD.  The present study indicates that the Wigner symmetry is realized even for the color number is three for the case of large pion mass, which corresponds to large quark mass.

On the other hand, in nature, the scattering lengths are $a_{^1S_0}=-23.740\pm 0.020$ fm and
{$a_{^3S_1}=5.419\pm 0.007$} fm for proton-neutron system~\cite{machleidt01,wang19}. Therefore, $\widetilde{C}_{^1S_0}$ and $\widetilde{C}_{^3S_1}$ should have distinct differences when the pion mass is close to the physical pion mass. 

\begin{figure}[htb] 
	\centering
	\includegraphics[scale=0.60]{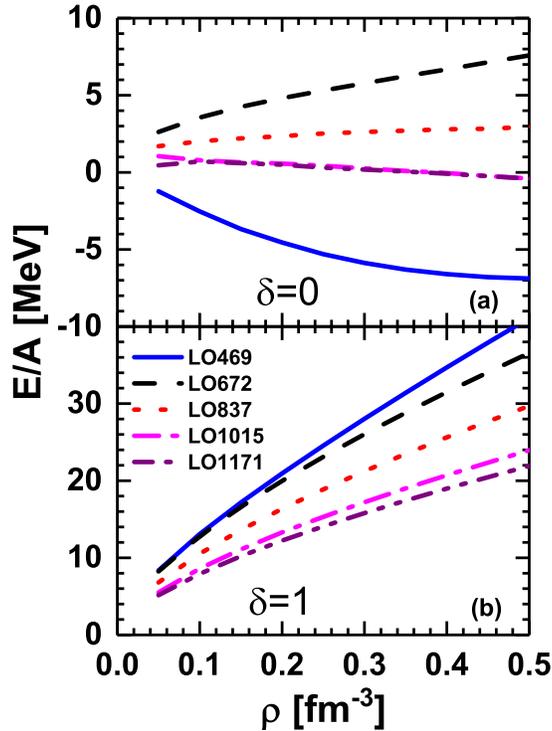}
	\caption{\label{eos} The equations of state of symmetric nuclear matter and pure neutron matter from LO chiral $NN$ potentials with different pion masses from $468.6$ MeV to $1170.9$ MeV. }
\end{figure}

The LO chiral $NN$ potentials from HALQCD data are adopted to investigate the properties of nuclear matter in the framework of BHF method. In these calculations, only four channels, $^1S_0,~^3S_1,~^3D_1$, and $^3S_1$-$^3D_1$ are included to follow the available LQCD simulations. In Fig.~\ref{eos}, the equations of state of symmetric nuclear matter (panel (a), $\delta=(N-Z)/A=0$) and pure neutron matter (panel (b), $\delta=(N-Z)/A=1$) are shown with different pion masses.  There is a saturation binding energy, $E/A=-6.80$ MeV at $\rho_B=0.45$ fm$^{-3}$ for LO469 potential in symmetric nuclear matter, while the bound system is not found in LO672 and LO837 potentials. With the pion mass, the systems have very weak bound states with LO1015 and LO1171 potentials at high density. Actually, the behaviors can be explained by the phase shifts of chiral potentials given in Fig.~\ref{ph}. The LO469 potential provides the largest phase shifts in the $^1S_0$ and $^3S_1$ channels, which generates the strongest attractive contributions. In LO672 potential, the phase shifts in the $^1S_0$ and $^3S_1$ channels suddenly decrease and are only one half of LO469 potential. With the pion mass, the magnitudes of phase shifts become larger and larger. Therefore, the energy per nucleon in the corresponding equations of state is changed from positive values to negative ones. Furthermore, the saturation binding energy of LO469 potential in present work is deeper than the similar calculation with HALQCD simulation potential directly by Inoue {\it et al.}, where the saturation point is $E/A=-5.4$ MeV at $\rho_B=0.414$ fm$^{-3}$~\cite{inoue13}. It is because that phase shifts of $^1S_0$ channel given by LO469 potential are more attractive than the lattice data at larger $E_\text{lab}$. In the pure neutron matter, only $^1S_0$ channel provides the contribution due to the Pauli principle.  The LO469 potential gives the most repulsive equation of state, while the LO1171 potential generates the most attractive one. In general, the behaviors of equations of state of nuclear matter in the LO chiral potentials are similar to those with HALQCD potentials obtained by Inoue {\it et al.}

\section{Conclusion}
Nature behaves as it is due to the fact that the pion mass is the one in the free space. The pion-mass dependence of $NN$ potentials was a very important problem to understand the QCD theory at the low energy scale.  The $NN$ potentials with pion-mass dependence were constructed {in term of one-pion exchange and contact terms}, for the first time, based on the chiral effective field theory with leading-order expansion and the recent lattice QCD achievements obtained by HALQCD group.  In these potentials, the attractions became weaker and weaker, and the LECs, $\widetilde C_{^1S_0}$ and $\widetilde C_{^3S_1}$ became larger with the pion mass. Furthermore, the magnitudes of $\widetilde C_{^1S_0}$ and $\widetilde C_{^3S_1}$ at larger pion masses were very closed to each other, while their differences became obvious when the pion mass approached the physical point. 

We constructed the equations of state of symmetric nuclear matter and pure neutron matter with these LO chiral potentials in the framework of BHF method. The LO469 potential generated the bound states in symmetric nuclear matter. As for pure neutron matter, there was only the contribution from $^1S_0$ channel due to the Pauli principle and the potentials with larger pion masses provided more repulsive contributions.  These results showed that the properties of nuclear many-body system were strongly dependent on the pion mass, which is a contradiction of the conclusions from the pionless ChEFT.  To interpret the relations between the lattice simulation and ChEFT on two-nucleon system, more partial waves are to be calculated in the lattice QCD, since next-to-leading order $NN$ interaction is combined by the LECs in terms of $^1S_0, ^3P_0, ^1P_1, ^3P_1, ^3S_1, ^3S_1-^3D_1$, and $^3P_2$ channels.

\section*{Acknowledgments}
	This work was supported by the National Natural Science Foundation of China (Grants  No. 11775119, No. 11675083, No. 11405090, and No. 11405116), the Natural Science Foundation of Tianjin, and China Scholarship Council (Grant No. 201906205013 and No. 201906255002).

\end{document}